\documentclass[%
superscriptaddress,
amsmath,amssymb,
aps,
twocolumn,
]{revtex4-2}
\usepackage{hyperref}
\hypersetup{colorlinks,allcolors=blue}
\usepackage{graphicx}
\usepackage{bm}
\usepackage[all]{hypcap} 
\usepackage{color}
\usepackage{siunitx}
\usepackage{mathtools}
\usepackage{ulem}

\begin{document}

\title{Controllable giant magneto resistance and perfect spin filtering in $\alpha^{\prime}$-borophene nanoribbons}

\author{Z. Askarpour}
\affiliation{\footnotesize{Department of Physics, Iran University of Science and Technology, Narmak, Tehran 16844, Iran}}

\author{M. Farokhnezhad}
\affiliation{\footnotesize{School of Nano Science, Institute for Research in Fundamental Sciences (IPM), Tehran 19395-5531, Iran}}

\author{F. Norouzi}
\affiliation{\footnotesize{Department of Physics, Iran University of Science and Technology, Narmak, Tehran 16844, Iran}}

\author{M. Esmaeilzadeh}
\email{mahdi@iust.ac.ir}
\affiliation{\footnotesize{Department of Physics, Iran University of Science and Technology, Narmak, Tehran 16844, Iran}}

\date{\today}
\begin{abstract}
By using non-equilibrium Green's function (NEGF) method and tight-binding (TB) approximation, we investigated a perfect control on spin transport in a zigzag $\alpha^{\prime}$-boron nanoribbon (zigzag $\alpha^{\prime}$-BNR) as the most stable semi-conducting structure of borophene. It has been found that, when an $\alpha^{\prime}$-BNR is exposed to an out-of-plane exchange magnetic field, spin filtering occurs for both spin-up and spin-down in specific ranges of energy, so that the spin polarization of current could be controlled by adjusting the energy of incoming electrons by means of an external back gate voltage. We focus on the edge manipulation of $\alpha^{\prime}$-BNR by ferromagnetic (FM) or anti ferromagnetic (AFM) exchange field which leads to the emergence of a giant magnetoresistance and a perfect spin filter. Local currents provide the best picture of spin distribution of current in the nanoribbon. In order to observe the response of the system to the proximity effect of magnetic strips, we calculated the magnetic moment of each site. Then, we show that applying a transverse or perpendicular electric field in the presence of the exchange magnetic field gives a another controlling tool on the spin polarization of current in a constant energy. Finally, by the simultaneous effect of in-plane and out-of-plane exchange magnetic field on the edges of nanoribbon, we reach a control even on the spin rotation in the scattering region. Our investigation guarantees the $\alpha^{\prime}$-BNR as a promising two dimensional (2D) structure for spintronic purposes.
\end{abstract}


\maketitle
\section{Introduction}

In the recent years, Borophene, a single layer of boron atoms, has attracted attention because of some specific features such as low density and high hardness , thermal resistance (melting point $2300 ^{o}C$ much higher than silicon) \cite{tian2008boron} and electronic conductivity \cite{liu2010metal}. Borophene, has been successfully synthesized on Ag, Cu, Ni and Au substrates \cite{large2019,borophene2019,experimental2018} and some of its outstanding properties such as novel magnetism \cite{magnetic2020}, electronic, optical, thermodynamic \cite{electronic2017,lopez,peng}, mechanical \cite{2015synthesis,mechanical2017,lattice2017} and superconducting properties \cite{penev}, have been investigated so far.

Four orbitals $S, P_x, P_y, P_z$ are available in borophene, but in comparison to graphene in which orbital $P_z$ is responsible for gapless bands, in borophene it mainly contributes to the bands near Fermi level based on first principle calculation and experiment \cite{dirac}. Boron atom is located between metals and non-metals in the periodic table of elements and resembles the most similarity to carbonwith a weaker spin-orbit coupling (SOC) and only one less valence electron which places it in different topological class in comparison to carbon and prevents it from forming graphene-like structures, while drives it to form complex and multicenter bonds. Consequently, different sort of boron-based structures have been found such as clusters, fullerene cage, nano tubes and 2D structures \cite{wu2012two,polymorphism,observation,synthesis2004} which are classified as $ \alpha, \beta, \chi, \psi, \delta $ nanostructures among which $ \alpha$, $\beta_{12}$ and $ \chi_{3}$ have been experimentally synthesized so far\cite{2015synthesis,feng} among which borophene nanoribbons, recently synthesized by Zhong \textit{et al.} \cite{zhong}, have been found to be the most suitable choice for electronic applications because of the high degree of electron confinement resembled as low transport resistance, size dependent band gap, edge dependent conductance and many other advantages.  
 
$\alpha$-boron, derived from Bucky ball $B_{80}$ structure \cite{novel,sadrzadeh}, as the lowest energy structure of borophene which has been predicted to show a high in-plane stiffness around four times of iron was supposed to be the most stable form of borophene \cite{peng2015}; however, it turns out, it is dynamically unstable against out-of-plane negative frequencies \cite{wu2012two} and undergoes a transformation to a slightly buckled shape in which every neighboring central atoms have shifted upward and downward of the plane. The new buckled structure named $\alpha^{\prime}$-boron (with the energy $E=-6.282$ which is even less than $\alpha$-boron) has been found to have the lowest energy among all borophene sheets \cite{semiconducting}. Contrary to $\alpha$-boron which was experimentally shown to be a metal as most phases of boron sheets, $\alpha^{\prime}$-boron was theoretically predicted to be a semiconductor and very recently it has been proved experimentally \cite{hou}. Therefore, $\alpha^{\prime}$-boron is the most stable semiconducting phase of borophene.

In both $\alpha$-boron and $\alpha^{\prime}$-boron, like graphene, orbital $P_z$ describes the states near Fermi level although in $\alpha^{\prime}$-boron, orbital $P_z$ is coupled to orbitals $P_x$ and $P_y$ because of the lack of mirror symmetry arising from the buckling and it is responsible for the metal-semiconductor phase transition from $\alpha$-boron to $\alpha^{\prime}$-boron. Since the conduction-band minimum (CBM) in $\alpha^{\prime}$-boron, like graphene, is dominated by $\pi$ bond, it has a high electron mobility (around 20,000 $ cm^{2} V^{-1} s^{-1} $). $\alpha^{\prime}$-boron, also shows a low resistance in contact with metallic phases of borophene sheets and makes it one of the best candidates to be used as a channel in all boron-based field effect transistors (FETs) \cite{semiconducting}. In spite of all advantages offered by $\alpha^{\prime}$-boron for electronic devices, very few works have been implemented in this regard because of its novelty.

 As mentioned before, borophene has a weak SOC and consequently a long electron coherence length which is an outstanding feature for spintronic purposes. Several works have been implemented in this regard such as spin transport in N-doped borophene \cite{sun}, borophene-$\chi_3$ and borophene-$\beta_{12}$ \cite{norouzi}. In this paper, by means of tight binding (TB) model and non-equilibrium Green’s function (NEGF) method, we investigate the charge and mainly spin transport properties of a zigzag $\alpha^{\prime}$-boron nanoribbon which has been exposed to out-of- plane exchange field in proximity to FM and AFM strips. Indeed, the exchange field removes the spin degeneracy of electrons and leads to spin splitting along the energy axis. Therefore, one can control the spin polarization of current by tuning the energy of incoming electrons by means of a back gate voltage. In order to trace the transport paths of both spin states and the magnetic distribution arising from the effect of magnetic strips, current distribution and magnetic moment have been studied. As a short explanation, the current distribution for spin-up and spin-down are rather different. The most noticeable difference is a region void of current in the center of nanoribbon for spin-up state and the most noticeable similarity is that the current is more likely to flow though the boundary of the hollow hexagons and these results are well in compliance with the local density of states (LDOS). Also, the magnetic moment which is calculated in a specific energy follows the spin density of states in this energy. Finally, a perfect control on spin splitting and spin rotation is attained by the simultaneous effect of out-of-plane and in-plane exchange fields. 

The rest of the paper is organized as follows: In Sec. \ref{sec2}, The main TB Hamiltonian in the presence of in-plane and out-of-plane magnetization and electric field along with a summary of Green’s function formalism and the local spin current, magnetic moment and spin polarization formalism. In Sec. \ref{sec3}, The results of our investigation have been explained in details. Finally, a conclusion of the work has been submitted in the Sec. \ref{sec4}.   

\section{MODEL AND FORMALISM}\label{sec2}
\subsection{Tight-binding Hamiltonian of $\alpha^{\prime}$-borophene}
 $\alpha^{\prime}$-borophene, a triangular lattice marked by hollow hexagons, with the buckled atoms in the center of filled hexagons (the vertical distance is around $0.188 {\AA}$) has the lattice constant a and b slightly less than $\alpha$-borophene. Due to the mirror asymmetry in $\alpha ^{\prime}$-borophene, $P_z $ state is coupled to $S$, $P_x$, $P_y$ orbitals which gives rise to opening a band gap in $ \Gamma $ point.
Based on the precise GW0 calculation by Zhang et al \cite{semiconducting}, the band structure of $\alpha^{\prime}$-borophene dominated by $P_z$, is described by an eight-band TB Hamiltonian as

\begin{eqnarray}\label{Eq.1}
H=\sum_{l,i,s}\varepsilon_{l,i,s}c^{\dagger}_{l,i,s}c_{l,i,s}+\sum_{{\langle{i,j}\rangle},s}tc^{\dagger}_{i,s}c_{j,s}\nonumber\\+\sum_{i,s}M_{z}c^{\dagger}_{i,s}\sigma_zc_{i,s}+\sum_{i,s,s^{\prime}}M_{x}c^{\dagger}_{i,s}{\sigma_x}c_{i,s^{\prime}} 
\nonumber\\+\sum_{i,s}ed_{i,s}E_{y}c^{\dagger}_{i,s}c_{i,s}+\sum_{i,s}e\Delta_{i,s}E_{z}c^{\dagger}_{i,s}c_{i,s},
\end{eqnarray}

\begin{figure}[t]
\centering
\includegraphics[scale=0.3]{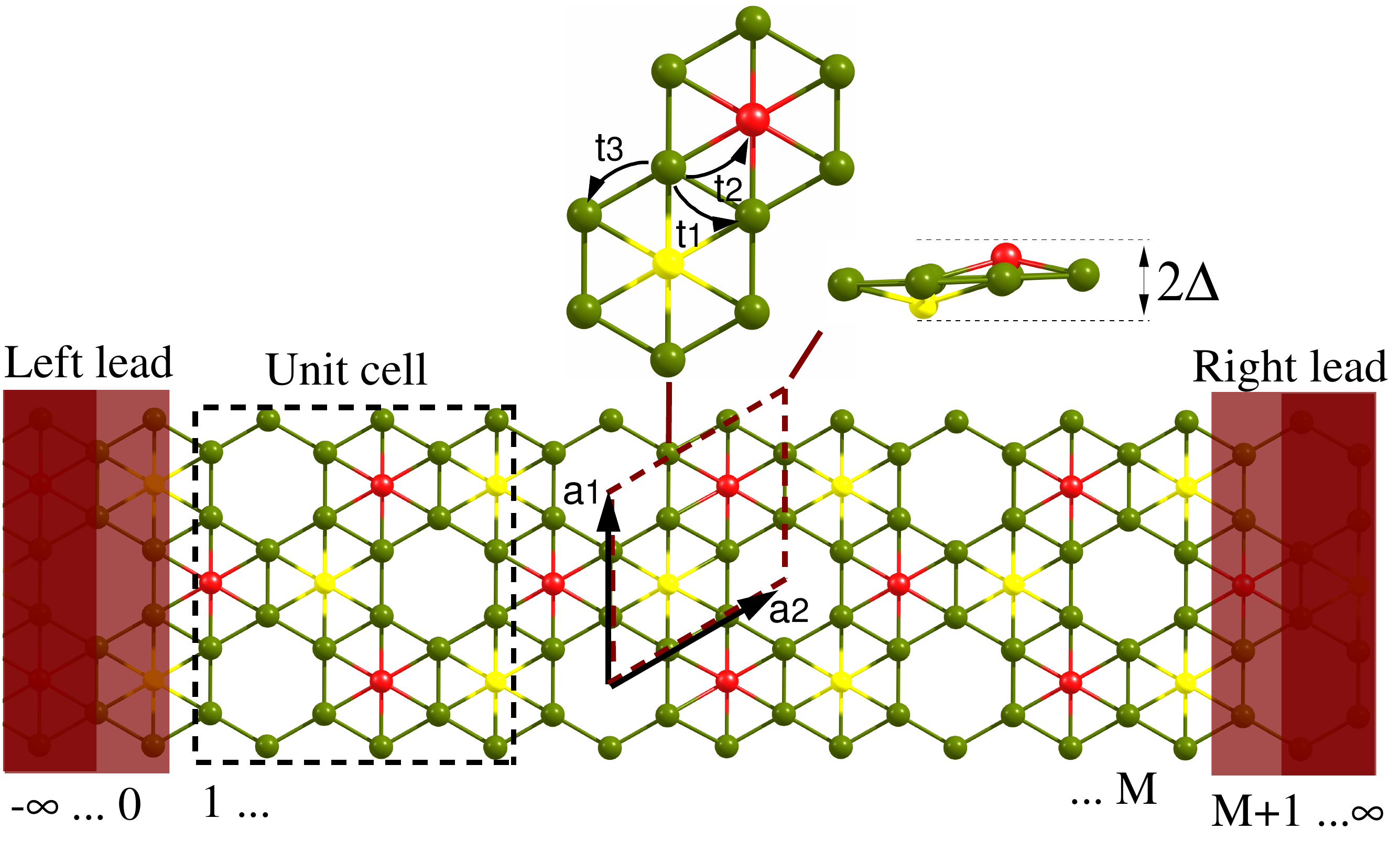}
\caption{\label{Fig1} Schematic view of a zigzag $\alpha^{\prime}$-BNR sandwiched between two semi-infinite leads from the same structure. The unit cell indicated by dashed wine parallelogram contains three kinds of atoms colored by green in the plane, red and yellow as buckled sites. The magnified view of the unit cell shows a better perspective of buckled sites and hopping energies between atoms. Here, $\textbf{a}_1=\ell (0,1)$ and $\textbf{a}_2=\ell(\sqrt{3}/2,1/2)$ are the lattice constants and $\ell=3a\simeq{1.68}$ {\AA} in which $a$ is the distance between two boron atoms. The scattering region includes $M$ super cells extended along the longitudinal direction and each super cell has 30 boron atoms in this paper.}
\end{figure}

In which, the operator $c^{\dagger}_{l,i,s}(c_{l,i,s})$ represents the electron creation (annihilation) at site $i$ of the sublattice $l$ with spin $s$ and $\langle{i,j}\rangle$ is limited to the nearest-neighbor sites.
Also, as shown in Fig.~\ref{Fig1}, $t_1$,$t_2$, and $t_3$ as the hopping energy between: the connecting atoms of two neighboring hollow hexagons, the central and marginal atoms of filled hexagons and two marginal atoms of hollow hexagons are:
\\$t_1=2.55$ eV    ,\:   $t_2=1.6$ eV    ,\:    $t_3=2.19$ eV\\

Ferthermore, $\varepsilon_1$ and $\varepsilon_2$ as the on-site energy of: the atoms in the plane of $\alpha^{\prime}$-BNR and the buckeled atoms in the center of filled hexagons are:
\\$\varepsilon_1=-1.9$ eV  ,\:   $\varepsilon_2=-2.2$ eV\\ 

These parameters are the same as that of $\alpha$-BNR except $t_2$ and $\varepsilon_2$ and this is because the $\pi$ bond between the central and marginal atoms of filled hexagons is affected by $S$, $P_{x}$ and $P_{y}$ orbitals. \cite{vishkayi} 
Furthermore, $M_{z}$ and $M_{x}$ are the out-of-plane and in-plane magnetizations which lie in different blocks of Hamiltonian by means of $\sigma_z$ and $\sigma_x$ as Pauli matrices and also $E_{y}$ and $E_{z}$ are the transverse and perpendicular electric fields which induce the potentials $ed_{i}E_{y}$ and $e\Delta_{i}E_{z}$ with $d_{i}$ and $\Delta_{i}$ as the distance from their origins respectively. It should be noted that, about the perpendicular electric field $E_{z}$, the potential distribution is as follows: the atoms in the plane of the lattice (green sites) in zero potential and the buckled atoms slightly over the plane (red sites) in $+e{\Delta}{E_z}$ potential and the buckled atoms slightly below the plane (yellow sites) in $-e{\Delta}{E_z}$ potential or vise versa for reverse direction of perpendicular electric field. 

Since the nano-ribbon can be considered as the periodic configuration of unit cells along the $x$ direction, one can write the $k_x$-dependent Hamiltonian based on Bloch theorem as

\begin{equation}\tag{2}
H(k_x)=H_te^{ik_{x}a}+H_0+H_te^{-ik_{x}a}
\end{equation}

Also, one can easily reach the energy dispersion of the band structure by diagonalizing this Hamiltonian.

in which, $H_0$ is the Hamiltonian of an arbiterary unit cell of the nano-ribbon, $H_t$ is the hopping energy between the unit cell with its adjacent unit cell, and $a$ is the lattice constant.

\subsection{Spin-dependent transport}

 In this stage, we use non-equilibrium Green's function to calculate the spin dependent conductance by using a famous iterative method proposed by Sancho \textit{et al.} which is applicable for a systemm with semi-infinite electrodes. It is worth to be noted that, to avoid inelastic scattering and phonon interaction, one should consider a suitable length of channel at low temperature. The super high mobility of $\alpha^{\prime}$-boron (21,148 $cm^{2}$V$^{-1}$s$^{-1}$ at 300 K) \cite{semiconducting} ensures us the condition for ballistic regime is satisfied even with a long channel.
The spin dependent conductance is calculated as 
 
\begin{equation}\tag{3}
G^{ss^{\prime}}(E)=\frac{e^{2}}{h}Tr[\Gamma_{L}^{s}(E)g^{ss^{\prime}}(E)\Gamma_{R}^{s^{\prime}}(E)g^{ss^{\prime}}(E)^{\dagger}],
\end{equation}

in which $G^{s,s^{\prime}}$ indicates the conductance of an incoming electron with spin $s$ and out going electron with spin $s^{\prime}$ and $E$ is the energy of the electron. It should be noted that only, in this paper, in the presence of in-plane magnetization, $s^{\prime}$ can be different from $s$; otherwise, they are the same. Also, $g^{s,s^{\prime}}(g^{s,s^{\prime}\dagger})$ represents the block $s,s^{\prime}$ of retarded (advanced) Green's function matrix

\begin{equation}\tag{4}
g(E)=[(E+i\eta)I-H_c-\Sigma_{L}-\Sigma_{R}]^{-1}
\end{equation}

in which, $\eta$ is an arbiterary infinitesimal number, $H_c$ is the Hamiltonian of the channel, and $\Sigma_{L}$ 

where, $\Sigma_{L(R)}$ is the self-energy of the left (right) lead which is given by:
 
\begin{equation}\tag{5}
\Sigma_{L}=H_{t}^{\dagger}g_{0}H_{t},
\end{equation}

in which $ H_{t} $ is the hopping matrix between two super cells of leads and $g_{0}^{s}$ is the surface Green's function which can be extracted by using the iterative method established by Sancho \textit{et al}\cite{sancho}.

Also, $ \Gamma_{L(R)} $ is the spin-dependent broadening matrix which describes the coupling between the channel and the left (right) lead of the system calculated by the spin-dependent version of the self-energy described above.

\begin{equation}\tag{6}
 \Gamma^{s}_{L(R)}(E)=i\left({\Sigma^{s}_{L(R)}-(\Sigma^{s}_{L(R)})^{\dagger}}\right)
\end{equation}

The spin-dependent local current at a specific energy $E$ between two neighboring sites $i$ and $j$ can be written as \cite{yan}
\begin{equation}\tag{7}
J_{i\rightarrow{j}}^{s}(E)=\frac{4e}{\hbar}Im[H_{ij}G_{ji}^{s,<}],
\end{equation}
where $G^{s,<}={g^s}{\Gamma^{s}_{L}}{g^{s\dagger}}$ is the spin-dependent lesser Green's function and $H_{ij}$ is the hopping energy between sites $i$ and $j$.

To observe the magnetic configuration as a response of the system to the proximity effect of an exchange field, one should calculate the magnetic moment of each site. 
\begin{figure}[t]
\centering
\includegraphics[scale=0.2]{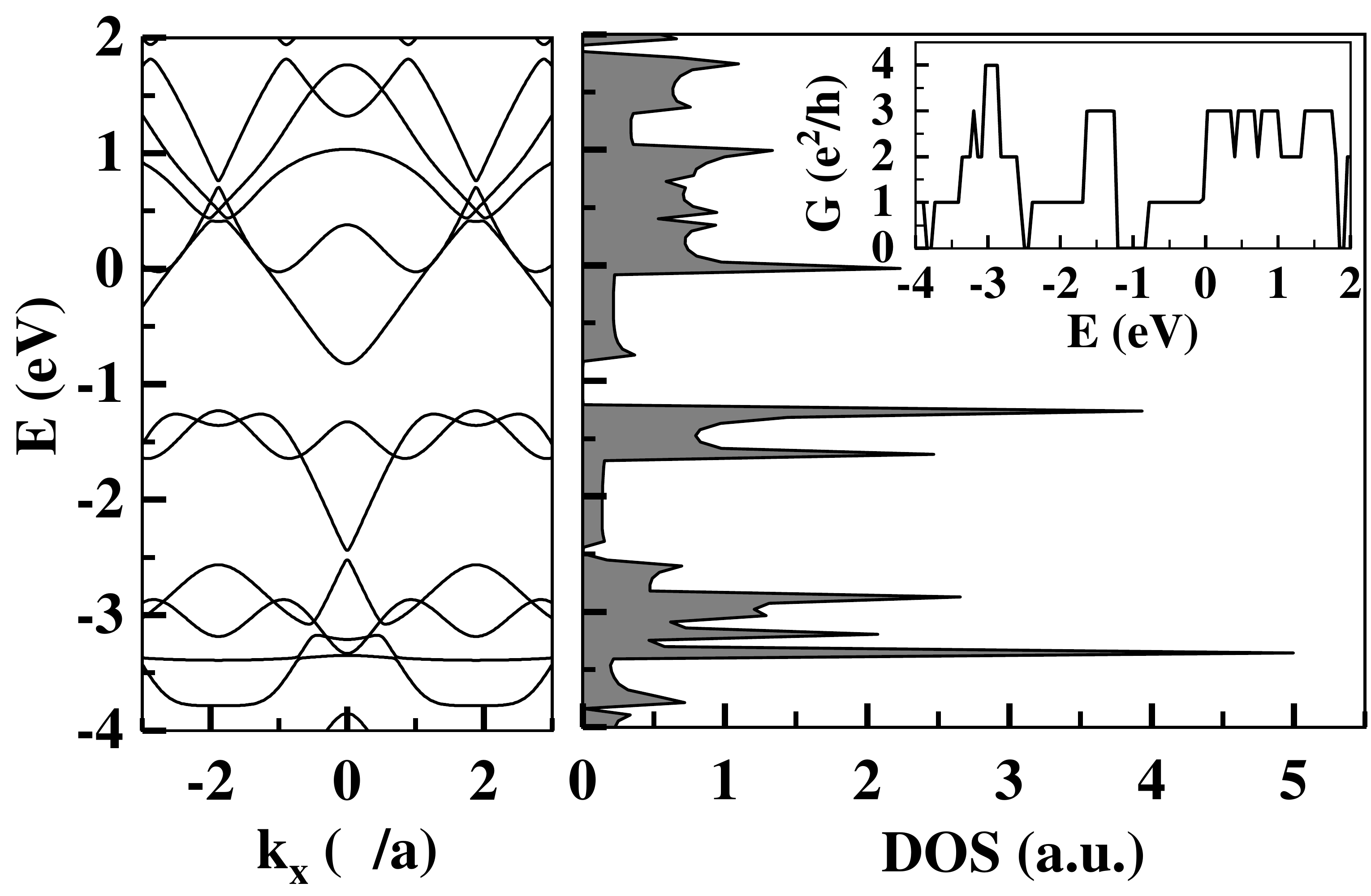}
\center
\caption{\label{Fig2}Band structure and density of states (DOS) and conductance in the inset of DOS for a zigzag $\alpha^{\prime}$-BNR introduced in Fig.~\ref{Fig1} which are completely in compliance with each other and introduce the $\alpha^{\prime}$-BNR as a semi-conductor with band gap $0.5024$ eV.}
\end{figure}

which is given by:
\begin{equation}\tag{8}
m_i=\mu_B\int_{-\infty}^{\infty}(\rho_{i}^{s}-\rho_{i}^{s^{\prime}})f(E-E_f)
\end{equation}
where, $\rho_{i}^{s(s^{\prime})}$ is the spin dependent local density of states at site $i$ and f($E-E_f$) is Fermi-Dirac function.

\begin{figure*}[t]
\centering
\includegraphics[scale=0.2]{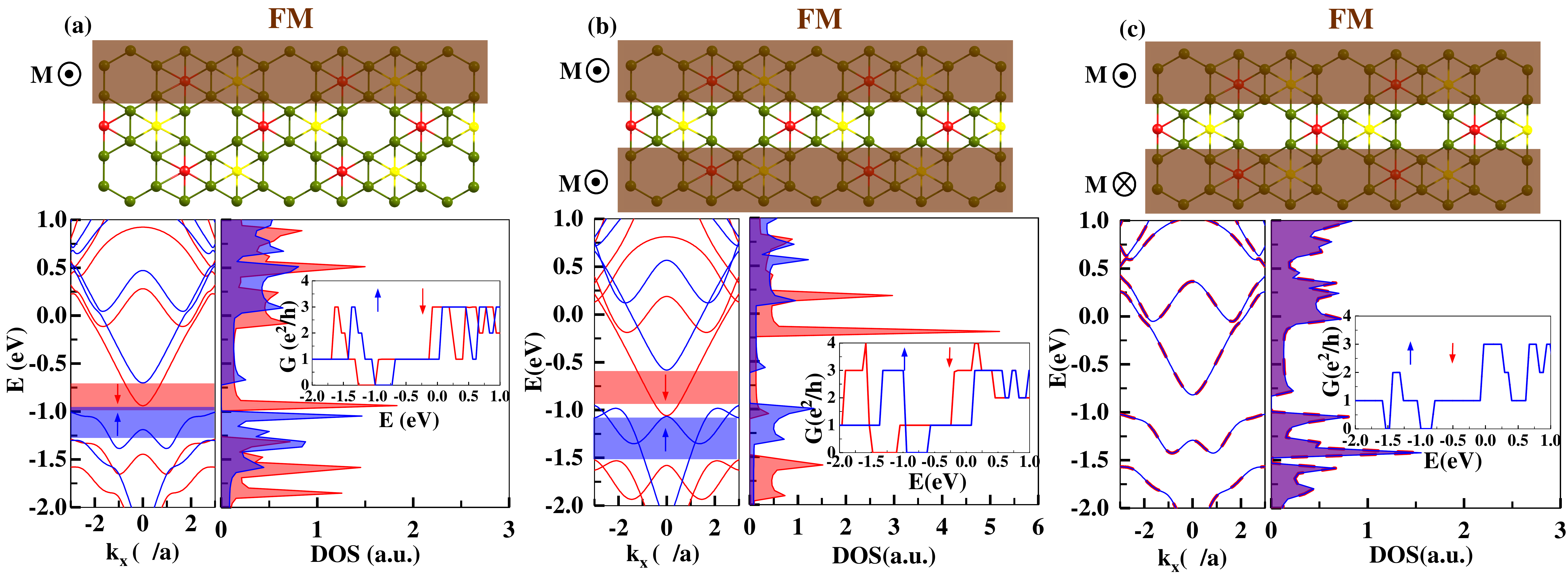}
\center
\caption{\label{Fig3}Spin spliting caused by non-local configurations of ferromagnetic strips exposed $\alpha^{\prime}$-BNR with the strength $M_z=0.3$ eV with (a) a single edge,(b) both edges with parallel ((c) antiparallel) configuration. Band structure and DOS with the Van Hove singularities corresponding to the energy sub bands in which the blue, red and purple (as the overlaping of blue and red) regions show the dominance (degeneracy) of spin-up, spin-down and (both spins), exhibit the spin spliting along energy axis. The insets of DOSs indicate the spin-dependent conductance as a function of the energy of the incoming electrons which conforms with the band structures and DOSs. The spin spliting in two edge exposed configuration is stronger than one-edge configuration as long as the magnetization caused by strips are parallel. In case of anti-parallel configuration no spin splitting occurs and both spin states are degenerate.}
\end{figure*}

The spin polarization of current could be calculated using spin dependent conductance\cite{farokhnezhad,jedema}

 \begin{equation}\tag{9}
P_{s}(E)=\frac{G^{\uparrow}-G^{\downarrow}}{G^{\uparrow}+G^{\downarrow}}. 
\end{equation}
$P_{s}$ can vary from -1 to +1. For negative (positive) value, it shows the conductance is dominanted with spindown (up) state.   

In the presence of in-plane magnetization $M_x$, spin polarization would have componants along x,y,z axes as

\begin{equation}\tag{11}
P_z=\dfrac{(G_{\upuparrows}+G_{\downarrow\uparrow}-G_{\downdownarrows}-G_{\uparrow\downarrow})}{G},
\end{equation} 

\begin{equation}\tag{12}
P_x=\dfrac{2e^2}{hG}Im(\sum_{nn^{\prime}\sigma}^Mt_{nn^{\prime}}^{*\sigma\uparrow}t_{nn^{\prime}}^{\sigma\downarrow})
\end{equation} 

\begin{equation}\tag{13}
P_y=\dfrac{2e^2}{hG}Re(\sum_{nn^{\prime}\sigma}^Mt_{nn^{\prime}}^{\sigma\uparrow}t_{nn^{\prime}}^{*\sigma\downarrow})
\end{equation}

Where M is the number of conducting channel and $\vert t_{n^{\prime},n,\downdownarrows}\vert^2$ is the probability of electron transmission with incoming spin $s$ in conducting channel n and outgoing spin $s^{\prime}$ in conducting channel $n^{\prime}$ which could be calculated from the below equation \cite{nikolic,controllable2015}:

\begin{equation}
\left(\begin{array}{cc}G^{\upuparrows} & G^{\uparrow\downarrow}\\ G^{\downarrow\uparrow}&G^{\downdownarrows} \\\end{array} \right)=\dfrac{e^2}{h}\sum_{n^{\prime},n}^M\left(\begin{array}{cc}\vert t_{n^{\prime},n,\upuparrows}\vert^2 &\vert t_{n^{\prime},n,\uparrow\downarrow}\vert^2\\\vert t_{n^{\prime},n,\downarrow\uparrow}\vert^2 & \vert t_{n^{\prime},n,\downdownarrows}\vert^2 \\\end{array}\right)
\label{pspec}
\end{equation}
                         
\section{RESULTS}\label{sec3}
In the following, we present our results calculated by employing TB Hamiltonian and NEGF method in a zigzag $\alpha^{\prime}$-BNR. It is composed of triangular lattice marked by isolated hollow hexagonal with $d_{3d}$ symmetry as shown in Fig.~\ref{Fig1}. The number of super cells along $x$ direction has been indicated by $M$ which each super cell includes $30$ atoms. Although we observed half-metallicity in all edge shapes of $\alpha^{\prime}$-BNR, in this research, we only focus on the zigzag shape because most of metallic phases of borophene which are suitable to be used in electrodes are zigzag , so using the same structure in the electrodes and channel only with different widths can reduce the mismatch effects in the system. First of all, the band structure and DOS and conductance shown in the inset of DOS with a small band gap assure us that we can consider $\alpha^{\prime}$-BNR as a semi-conducting phase of borophene. The steep energy bands are the indicator of high mobility and the bands near the Fermi energy are mostly localized on the edges of the nanoribbon which introduce the topological behavior in this material. 

\begin{figure*}[t]
\centering
\includegraphics[scale=0.2]{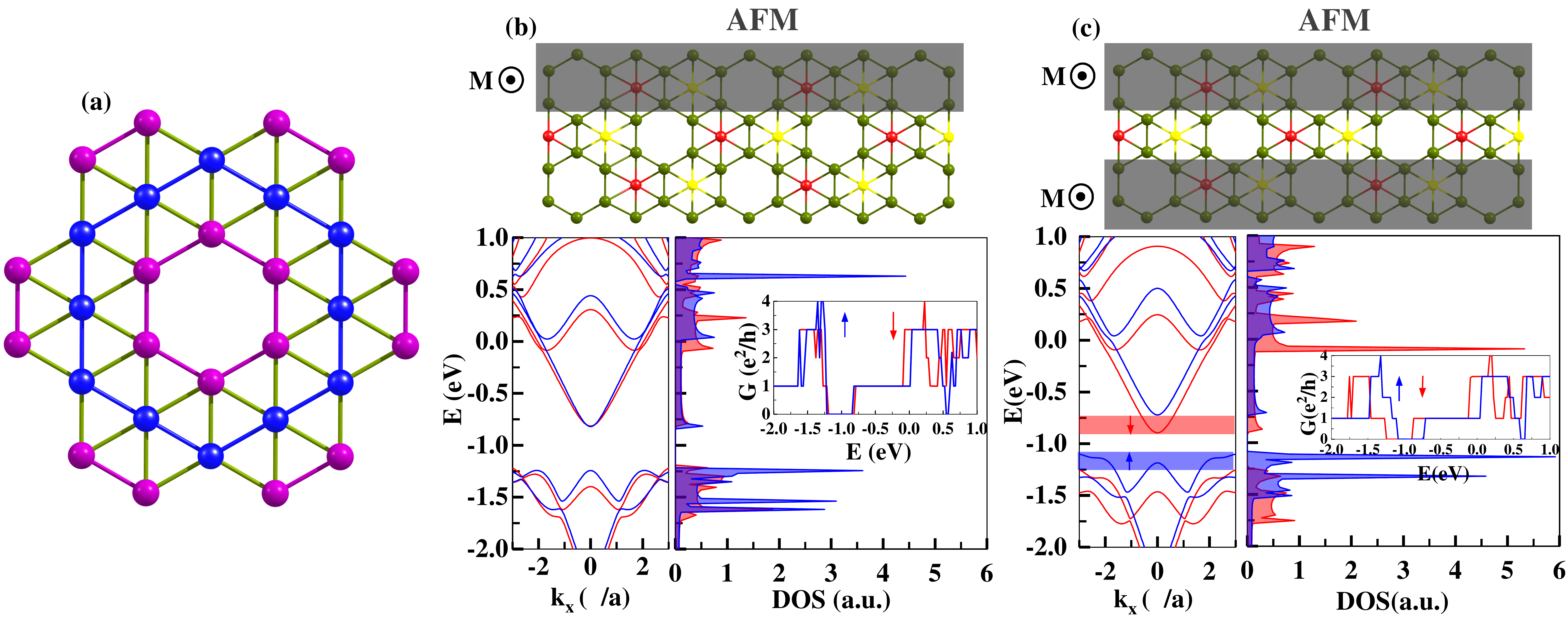}
\caption{\label{Fig4}(a)  a schematic view of magnetic distribution by a strip-type AFM strip. Spin spliting caused by non-local configurations of anti-ferromagnetic (AFM) strips exposed $\alpha^{\prime}$-BNR with the strength $M_z=0.3$ eV with (a) a single edge,(b) both edges configuration. Band structure and DOS with the Van Hove singularities corresponding to the energy sub bands in which the blue, red and purple (as the overlaping of blue and red) regions show the dominance (degeneracy) of spin-up, spin-down and (both spins), exhibit the spin spliting along energy axis. The insets of DOSs indicate the spin-dependent conductance as a function of the energy of the incoming electrons which conforms with the band structures and DOSs. The spin spliting in two edge exposed configuration is stronger than one-edge configuration.}
\end{figure*} 

 For the sake of our aim, we need to break the spin symmetry in the system. Inducing an exchange field in the nanoribbon via the proximity effect arising from a  ferromagnetic or anti-ferromagnetic substrate or strip is the most feasible way for this purpose. Contrary to other techniques for inducing magnetization in the system such as doping with transition metal atoms \cite{stability2013,chang2013} or bulking \cite{newly}, this approach offers the most convenient way of external manipulation with controllable penetration depth in which electron mobility is not affected which makes it the best candidate for electronic applications.
Let us start with the non-local ferromagnetic case which extends through the system consists of channel and leads. It must be noted that all magnetic manipulations in this paper are in the edge of nanoribbon because first, it is more controllable to involve as many sites as we need, moreover, it doesn’t destroy the topological properties of the system.

To move forward, only one ferromagnetic strip is deposited on one edge of the nanoribbon. No matter how many sites get involved, spin splitting occurs as soon as magnetization in induced on sites although it is stronger for wider strips when more sites are involved. In our work, each strip covers 5 sites along y axis, in this step with the out-of-plane exchange field $M_{z}$. As shown in Fig.~\ref{Fig3} the spin up and down bands split and shift toward different direction along energy (E) axis and the resultant spin gap is directly proportional to the strength of the exchange field ($\sim 2M_z$). This spin polarization could be traced in DOS figure marked by spin polarized Van Hove singularities. One can practically determine the spin of incoming electron by tuning its energy by means of an external back gate voltage. To be more precise, as shown in the inset of Fig.~\ref{Fig3} (a), the conductance of the system is completely spin polarized in the range of $-1.25<E<-1$ eV ($-1<E<-0.75$ eV) , so only spin up (down) can passes through the nanoribbon in these energies. The opposite could happen only by reversing the direction of magnetization.  

Finally, it could be concluded that the $\alpha^{\prime}$-BNR can be considered as a half-metal since it shows metalic behavior for one spin state and insulating for the other one. In fact, it acts as a controllable spin filter system as if in a specific range of energies, it plays the role of an insulator for one spin state and a metal for the other spin state at the same time or conversely for another range of energy \cite{yazyev}. 
To go further, now we deposit another ferromagnetic strip on the other edge of nanoribbon. Depending on the parallel or anti parallel configuration of the strips, two far different behavior could be observed in the system. As seen in Fig.~\ref{Fig3} b spin splitting occurs in $-1.5<E<-1.1$ eV ($-0.9<E<-0.6$ eV) and as expected, it is stronger than the case with one edge involved, while spin degeneracy is preserved and no spin splitting occurs for anti-parallel configuration as shown in Fig.~\ref{Fig3}c. 

For more investigation, we examine the anti-ferromagnetic effects this time. Because of the existence of Br-atoms in the center of some hexagons, anti-ferromagnetic distribution in borophene is challenging contrary to graphene and silicene in which the next nearest neighbors simply have the same magnetization sign. Using an strip-type anti-ferromagnetic strip resolves this problem completely. To imagine the anti-ferromagnetic distribution in this way, one can consider a hollow hexagonal in $\alpha^{\prime}$-boron lattice and change the sign of magnetization with each jump over the concentric bigger hexagons as shown in Fig.~\ref{Fig4}(a) as if purple sites have $M_z$ and blue sites have $-M_z$ magnetization or vice versa.

Like the ferromagnetic case, exposing both edges of $\alpha^{\prime}$-BNR with AFM strips exhibits a better spin splitting rather than the case in which a single edge is involved as shown in Figs.~\ref{Fig4}(b),(c) although either of them shows a weaker spin splitting in comparison to when we deposited the FM strips.

\begin{figure}[t]
\centering
\includegraphics[scale=0.4]{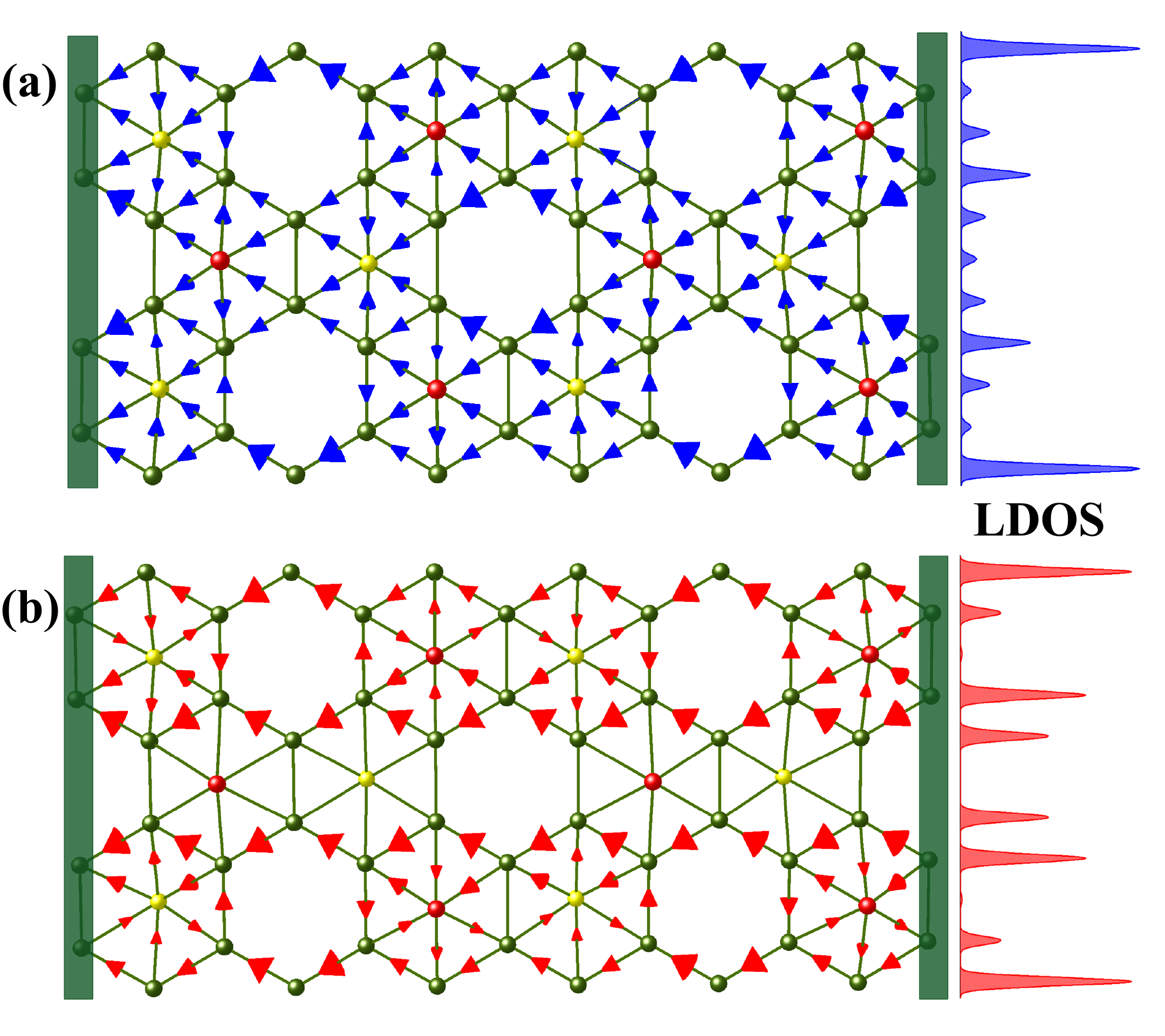}
\caption{\label{Fig5} Local current distribution and LDOS in the presence of parallel FM strips with the strength $M_z=0.3$ eV for (a) Spin-up states with the energy of incoming electrons $E=-1.4$ eV and (b) Spin-down states with the energy of incoming electrons $E=-0.8$ eV. The size of arrows is the indicator of local current magnitude flowing between two neighboring sites.}
\end{figure}
Now that we achieved a perfect spin polarized current, we are intended to trace the path of spin-up and spin-down states in the lattice. For this purpose, we study the distribution of local spin polarized current, shown in Fig.~\ref{Fig4}, in the presence of, for example, parallel FM strips as presented in Fig.~\ref{Fig3}b. First of all, we need to set the energy of incoming electrons as $E = -1.4 (-0.8$ eV) to observe the path of fully spin-up (down) polarized current respectively. Interestingly, we see nearly different, although with some similarities, current distribution for spin-up and spin-down states. Here, the right lead serves as the source and the size of arrows is the indicator of local current magnitude flowing between two neighboring sites. As shown in Fig.~\ref{Fig5}(a) local current with spin-up state is spread over all sites although with different proportions, while Fig.~\ref{Fig5}(b) shows a region void of current in the center of the nanoribbon for spin down state and also one can see that the most of current flow is passing through the boundaries of this region. It resembles a good similarity to edge modes in topological insulators. To reach a better comprehension, LDOS for both spin states are shown in the right side of the figure in which each peak is the  LDOS summation of horizontally aligned sites and the summation of two neighboring peaks is proportional to the local current passing between corresponded neighboring sites. It helps us to explain why the most of current flow is passing through the boundaries of hexagons, while a weak current flow is passing through the central atoms (red and yellow sites) and this is because LDOSs for sites in the boundary (center) of hexagons are maximum (minimum), and this difference is stronger for spin-down states.

\begin{figure}[ht]
\centering
\includegraphics[scale=0.4]{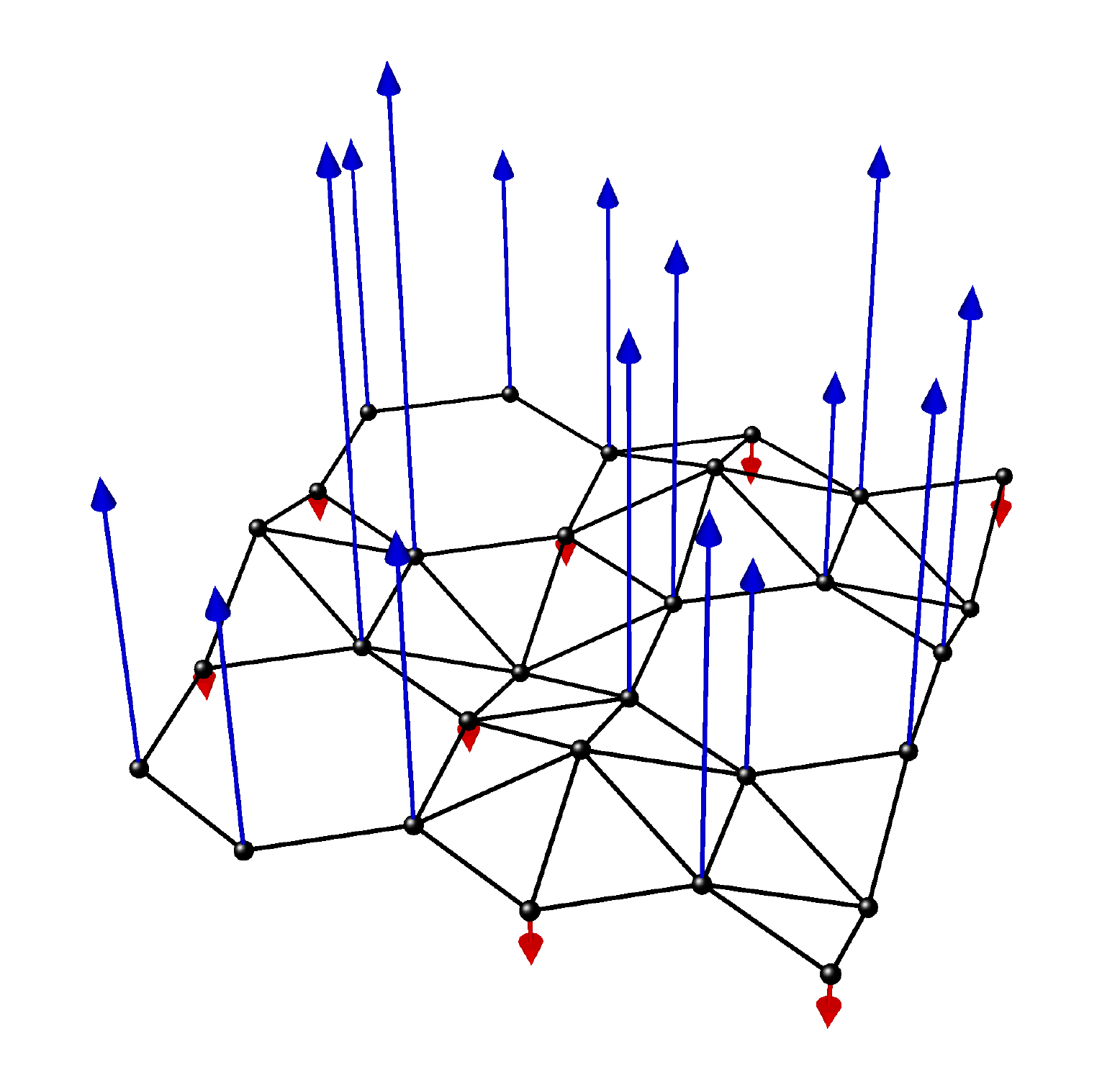}
\caption{\label{Fig6} Magnetic moment of a zigzag $\alpha^{\prime}$-BNR supercell with configuration as Fig.~\ref{Fig3}(b) in energy E=-1 eV. The magnetic moment for majority (minority) spin is illustrated by blue (red) arrows.}
\end{figure}  

 Finally, what is completely alike in both spin states is the current passing through the upper and lower edges of the nanoribbon and also the vertical boundary between hexagons and more interestigly, the topological behavior of the system which is an outstanding feature for this material. 

Through what have been studied so far, we have observed a good spin splitting arising from induced magnetization by magnetic strips. To go further, we are intended to implement a more fundamental investigation. Indeed, in order to observe the magnetic response of the system which is the root cause of this phenomenon,  we study the magnetic moment in this stage. For this purpose we need to calculate the LDOS of spin-up as the majority spin and spin-down as the minority spin in a specific energy (i.e. E=-1 eV in Fig.~\ref{Fig3}(b)) in which the conductance of spin-up is more than spin-down that necessarily means there are more channels available for the transport of spin-up state in this energy. Fig.~\ref{Fig6} shows the magnetic moment of configuration introduced in Fig.~\ref{Fig3}(b), for majority (minority) spin by blue (red) arrows and one can easily see the total magnetic moment follows the conductance.            

 We are intended to find another parameter other than the energy of incoming electrons by which the spin polarization of current can be controlled and applying a transverse or perpendicular electric field is found to do so. 

Up to now, we have investigated the effect of nonlocal exchange magnetic field applying on the edges of nanoribbon, while access to the entire system by gate and substrate, in practice,  may not be always possible. In order to avoid the drastic mismatch effects between channel and leads, in this stage, we confine the effect of electric field only to the channel, but the magnetic strips still are extended all over the edges of nanoribbon considering both channel and leads although in the following we investigate the case in which magnetic strips only exist in the channel.

\begin{figure}[t]
\centering
\includegraphics[scale=0.2]{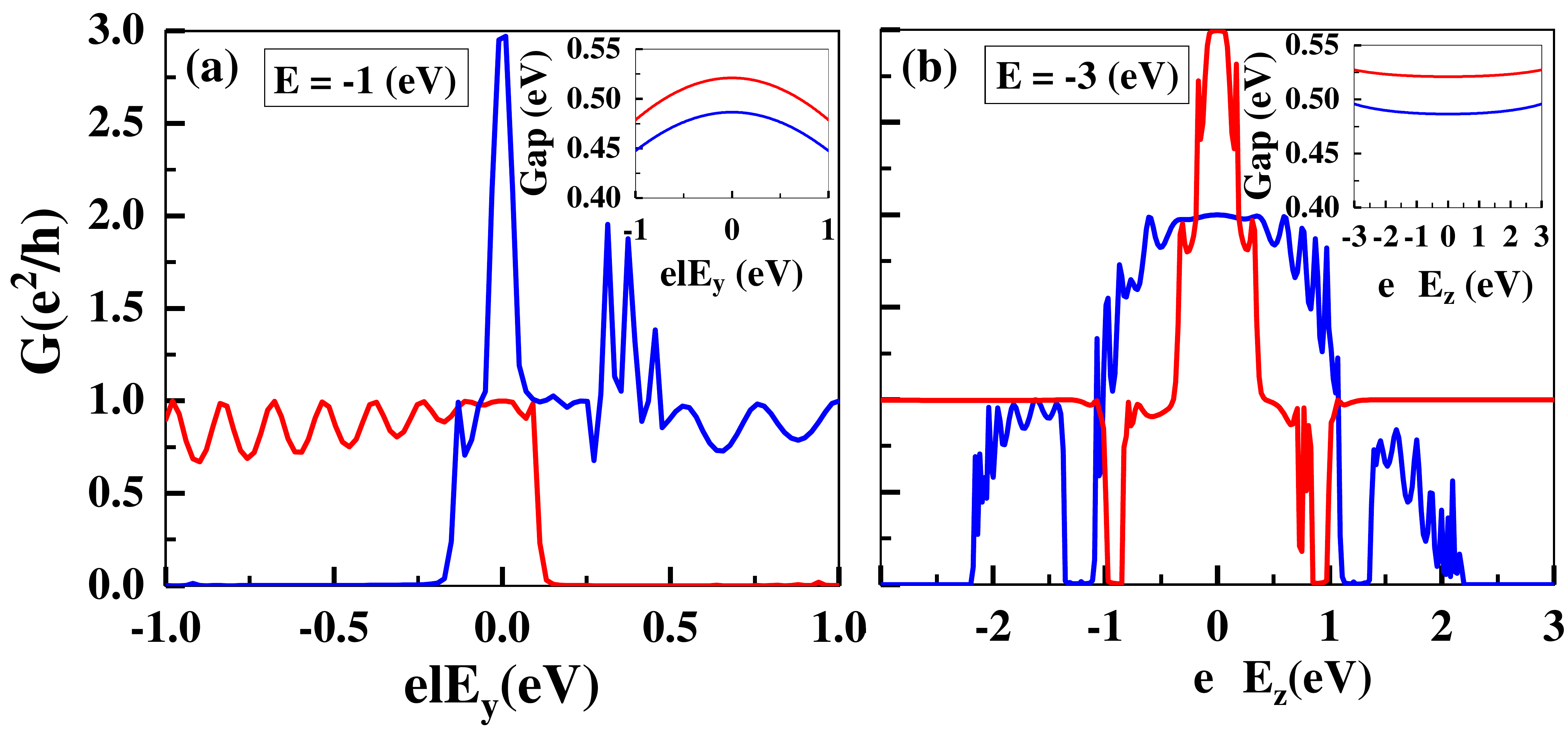}
\caption{\label{Fig7}Spin-dependent conductance for a nan-local parallel FM exposed $\alpha^{\prime}$-BNR with the strength $M_z=0.3$eV with inset figure of spin-gap size as functions of potential induced by (a) the transverse electric field $e{\ell}{E_y}$ applied on the channel for a specific energy ($E=-1$ eV) of incoming electrons and (b) the perpendicular electric field $e{\Delta}{E_z}$ applied on the channel for a specific energy ($E=-3$ eV) of incoming electrons. The figures shown in the insets indicate the size of spin band gap as a function of the electric fields. As the transverse (perpendicular) electric field increases in either direction, the band gap for both spin states decrease (increase). This behavior is steep (mild) in the case of transverse (perpendicular) electric field.}
\end{figure}

Now let us start with the transverse electric field which induces a gradient of potential over sites along $\hat{y}$ axis. Again, we consider the parallel configuration of FM strips in Fig. \ref{Fig3}(b) with the same strength $M_z=0.3$ eV, for instance, and apply the transverse electric field only on the channel of the device. As a result of this calculation, the spin-dependent conductance as a function of the potential induced by the transverse electric field $E_y$ for a specific energy ($e.g. E=-1$ eV) of incoming electrons is illustrated in Fig. \ref{Fig7}(a) which shows a zero (non-zero) conductance for the electrons with spin up (down) state in the range of ($-1$ eV $<e{\ell}{E_y}<-0.2$ eV) in case the electric field is in $-\hat{y}$ direction, and a zero (non-zero) conductance for the electrons with spin down (up) state in the range of ($0.15$ eV $<e{\ell}{E_y}<1$ eV) in case the electric field is in $+\hat{y}$ direction, so that the system can act as a perfect spin filter and one can change the spin polarization of current only by reversing the direction of a transverse electric field in a specific energy. This outstanding characteristic enables $\alpha^{\prime}$-BNR to be applicable in spintronic nano-devices. 

Now it is the perpendicular electric field turn. Considering the same configuration (the parallel FM strips over the nanoribbon as shown in Fig. \ref{Fig3}b with the same strength $M_z=0.3$ eV) as that of the previous case, this time, we apply a perpendicular electric field on the channel. We assume that the potential distribution as follows: the atoms in the plane of the lattice (green sites) in zero potential and the buckled atoms slightly over the plane (red sites) in $+e{\Delta}{E_z} $ potential and the buckled atoms slightly below the plane (yellow sites) in $-e{\Delta}{E_z} $  potential or vise versa for reverse direction of perpendicular electric field.
Now we fix the energy of incoming electrons at $E=-3$ eV where the LDOS of buckled atoms is mainly localized. As seen in Fig. \ref{Fig7}(b) although reversing the direction of the perpendicular electric field makes no difference because of the high degree of symmetry in the lattice, the spin-dependent conductance changes its domination for different values of induced potential for either direction many times; therefore, one can control the spin polarization of current in a specific energy of incoming electrons by adjusting the strength of a perpendicular electric field.

It is worth to remind that the fluctuation in both figures is due to the mismatch effect because of applying electric field only on the scattering region, while if it had been applied all over the system, the conductance would have showed a smoother behavior. 

In addition, the size of spin-gap as a function of potential induced by electric field has been illusterated in the inset of this figure based on which one can see a similar behavior for both spin states while they show decreasing (increasing) behavior against transverse (perpendicular) electric field. What is notable here is that the spin-gap is tunable by both electric field. 
    
\begin{figure}[t]
\centering
\includegraphics[scale=0.16]{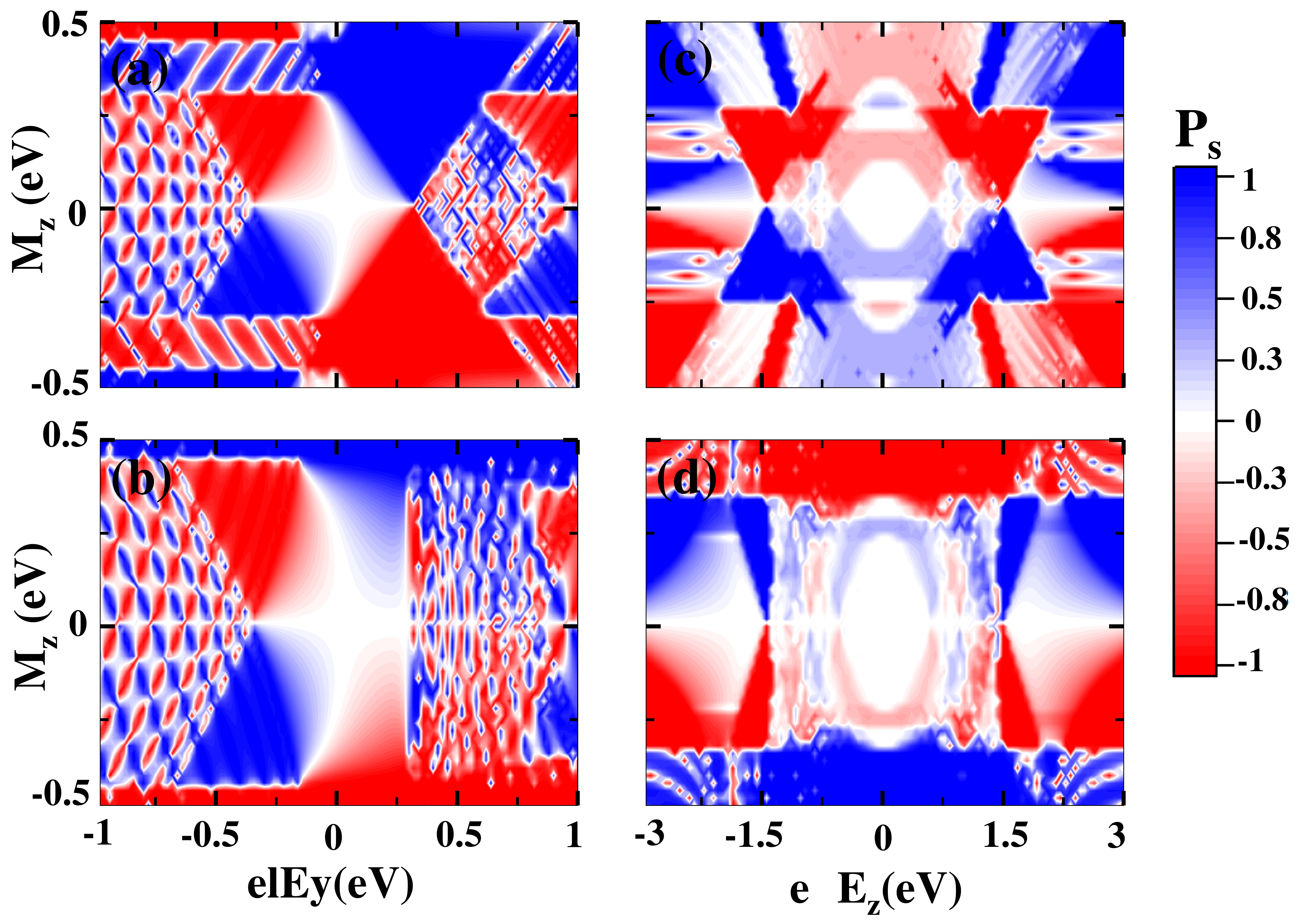}
\caption{\label{Fig8} Contour plot of the spin polarization $P_{s}$ in a constant energy ($E=-3$ eV) as a function of the exchange field $M_z$ and the potential induced by the transverse (perpendicular) electric field $E_y$ ($E_z$) for the configuration with two edges exposed by (a,c) FM strips (b,d) AFM strips. Blue (red) region shows the perfect spin up (down) polarization. The pale colors show a degree of spin polarization and white region is the indicator of spin degeneracy of states or simply a gap in energy bands.}                     
\end{figure}

In order to achieve a complete model for showing the influence of different strengths and directions of magnetization and electric field on the system, a contour plot of spin polarization $P_s$ for two edges exposed to FM and AFM strips (Fig.~\ref{Fig3}(b)and Fig.~\ref{Fig4}(b)) as a function of the exchange field $M_z$ and the potential induced by the transverse and perpendicular electric field is illustrated in Fig.~\ref{Fig8} in which the region marked by blue (red) shows a perfect spin polarization of spin-up (down) state and in the pale blue and red regions we have a degree of spin dominancy in the favor of spin-up (down) state, and finally the white region indicates the degeneracy of both spin states. The regions with swinging patterns indicate the states in which the systemis not stable enough and change its polarization with a small change in the strength of magnetization and electric field, so one had better set the parameters in the regions with a smooth pattern to be benefited by the spin polarization of the system. As seen in this figure, the perfect spin polarization is achievable for both configurations in the presence of the perpendicular electric field but obviously FM strips can give a better spin polarization for our system, in the presence of the transverse electric field. It is worth to be noted that, the pattern of the spin polarization isn’t preserved by reversing the direction of the transverse electric field, while reversing the direction of the perpendicular electric field makes no difference and it is due to the out-of-plane symmetry of the lattice.
 
To go further, we are willing to survey the conservation of spin in the system. As mentioned before, applying in-plane exchange magnetic field causes the spin polarized Hamiltonian not to commute with the spin operator $S$. It implies that the spin state may changes along the transport direction. To be more precise, when an electron enters the channel, one may expect that its spin state, spin up for example, persists its motion along the transport direction and leaves the channel toward the drain still as spin-up state, while this scenario isn't true when in-plane exchange magnetic field exists. In fact, as a result of being exposed to in-plane exchange magnetic field, spin rotation occurs and spin-up state may leave the channel as spin-down state. 

\begin{figure}[t]
\centering
\includegraphics[scale=0.19]{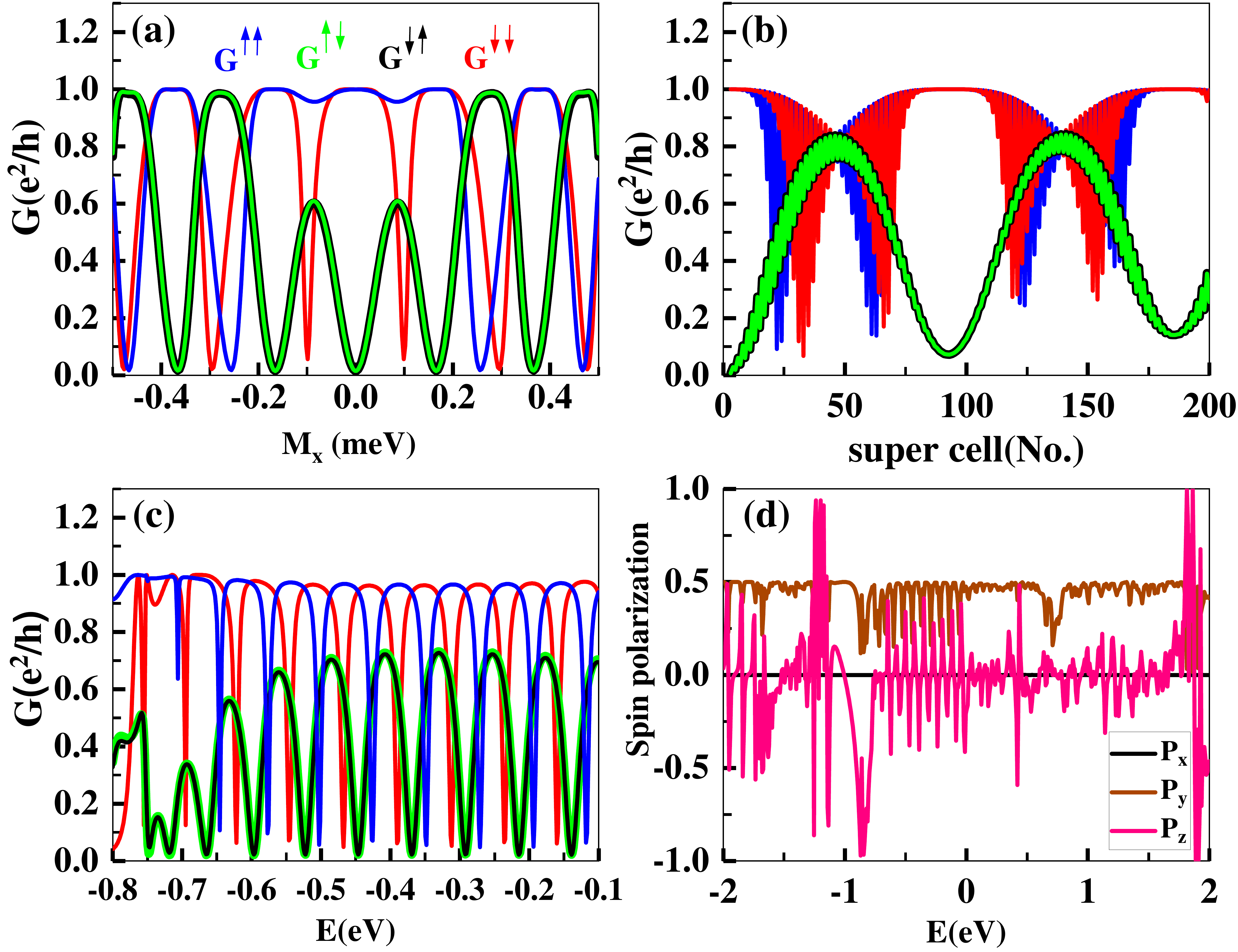}
\caption{\label{Fig9} Spin dependent conductance as a function of (a) in-plane exchange magnetic field in constant energy ($E=-0.38$), (b) the number of supercells in the channel in constant energy ($E=-0.19$), (c) energy of incoming electrons for configuration with one edge exposed by $M_z=0.1$ eV and the other edge by variable $M_x$. (d) all componants of the spin polarization as a function of energy. The The first (last) arrow in $G^{\downarrow\uparrow}$, for example, shows the spin of the electron entering (leaving) the scattering region.}                     
\end{figure}

In this stage we confine the effect of magnetic strips only to the channel (i.e. local case) . For the sake of completeness, we place an out-of-plane FM magnetic strip over one edge of the nanoribbon and an in-pane FM strip over the other edge. It provides us with the opportunity to study the spin splitting and spin conservation in the system simultaneously. As illustrated in Fig.~\ref{Fig9}(a), both characteristic for both spin states are shown as a function of in-plane exchange magnetic field $M_x$ for a constant value of out-of-plane exchange magnetic field $M_z=0.1$ eV and energy of incoming electrons $E=-0.38$ eV. As mentioned before, the effect of $M_z$ causes the spin splitting occurs once the current flow enters the channel. The conductance for spin-up (down) state is plotted, as usual, by blue $G^{\uparrow\uparrow}$ (red$G^{\downarrow\downarrow}$) curve respectively. Particularly, one of these arrows describes the spin state when the electron leaves the source and the other describes the spin state when it enters the drain, so these curves belong to the case in which the spin-states are the same when electron enters and leaves the channel although they may change along the transport direction, while there are other curves as $G^{\uparrow\downarrow}$ and $G^{\downarrow\uparrow}$ in the figure which describe the spin rotation due to the effect of in-plane magnetic field.

This characteristic could be highly applicable in electronic devices since one can determine the spin polarization of output current completely different from the spin polarization of input current. Consider two specific values $M_x=0.1 (0.25)$ eV in Fig.~\ref{Fig9}(a) in which both spin states are allowed to enter the scattering region, while only spin-up (down) state is preserved along the transport direction and spin down (up) is rotated inevitably. This story also depends on the length of scattering region, as shown in Fig.~\ref{Fig9}(b), spin rotation occurs for the length of scattering region corresponding to $\thickapprox$ 50 and 150 supercells while spin is rather conserved for $\thickapprox$ 180, 90 and less than 25 cells. Since the energy of electrons could be easier changed than the strength of magnetic strips, the spin-dependent conductance as a function of energy in a specific strength of the exchange field ($M_x,z=0.1$ eV)is calculated as Fig.~\ref{Fig9}(c) and fortunately the same behavior could be observed again. Finally, all components of the spin polarization as a function of energy is illustrated in Fig.~\ref{Fig9}(d) in which the spin polarization along x ($P_x$) axis is zero, while the other components ($P_y,P_z$) are in compliance with the conductance curve in the related range of energy.
     
\section{SUMMARY}\label{sec4}

In this paper, we investigated the spin transport properties in an $\alpha^{\prime}$-BNR with zigzag edge shape in the presence of exchange magnetic field through a proximity to a ferromagnetic and anti-ferromagnetic strips. One can achieve a spin polarized current by tuning the energy of incoming electrons by means of a backgate voltage. It has been found that applying a transverse or perpendicular electric field can give a control on the spin polarization of current even in a specific energy. The spin distribution of current and the magnetic response of the matter have been studied through local current and magnetic moment respectively. Finally we achieved a control on the spin rotation through the transport direction by applying in-plane exchange magnetic field. Our results guarantee the $\alpha^{\prime}$-BNR as a promising candidate for spintronic application. 

\bibliographystyle{apsrev4-2} 
\bibliography{aps} 

\end{document}